\documentclass[12pt]{article}

\usepackage{amsmath,amsfonts,amssymb}

\def\bA{\mathbf{A}}

\def\be{\begin{equation}}

\def\ee{\end{equation}}

\def\bea{\begin{eqnarray}}

\def\eea{\end{eqnarray}}

\def\mH{\mathcal{H}}

\def \bA{\mathbf{A}}

\newcommand{\mL}{\mathcal{L}}

\begin{document}

\begin{titlepage}

\vskip 0.4 cm

\begin{center}
{\Large{ \bf Note About  Covariant Hamiltonian Formalism for Strings, p-Branes and Unstable  Dp-Branes
}}

\vspace{1em}  Josef Kluso\v{n}$\,^1$
\footnote{Email address:
 klu@physics.muni.cz}\\
\vspace{1em} $^1$\textit{Department of Theoretical Physics and
Astrophysics, Faculty
of Science,\\
Masaryk University, Kotl\'a\v{r}sk\'a 2, 611 37, Brno, Czech Republic}\\

%
%

\vskip 0.8cm

\end{center}

\begin{abstract}
In this short note we  formulate Covariant Hamiltonian formalism for
strings, p-branes and Non-BPS Dp-branes. We also  analyse the vacuum
tachyon condensation in case of unstable D1-brane.

\end{abstract}

\bigskip

\end{titlepage}

\newpage

\section{Introduction and Summary}
Actions for extended objects in string theory can be considered as specific form
of relativistic theories defined on the world-sheet (in case of fundamental string) or
world-volume (in case of p-branes, Dp-branes)
\footnote{For review, see for example \cite{Polchinski:1998rq,Polchinski:1998rr,Ortin:2015hya}.}. When we analyse dynamics
of these objects we usually derive Euler-Lagrange equations of motions. On the other hand it turns out that in some situations it is more convenient to switch to Hamiltonian formalism. One such a famous example is an analysis of the fate of the unstable Dp-brane at the tachyon vacuum \cite{Sen:2004nf} since Lagrangian for unstable Dp-brane is multiplied by tachyon potential \cite{Sen:1999md,Bergshoeff:2000dq,Kluson:2000iy}
 which means that it vanishes at the tachyon vacuum. On the other hand  corresponding Hamiltonian density is finite at the tachyon vacuum and describes dynamics of the gas of fundamental strings
\cite{Sen:2000kd,Sen:2003bc}. However main drawback of the Hamiltonian formalism is the lack of the manifest diffeomorphism invariance since we have to select time coordinate
as special one. On the other hand there exists covariant Hamiltonian formulation of field theory known as Weyl-De Donder theory
\cite{DeDonder,Weyl}. The key point of this formulation is that now canonical Hamiltonian
density depends on conjugate momenta $p^\alpha_M$ which are variables conjugate to $\partial_\alpha x^M$. In other words we tread all partial derivatives on the equal footing which clearly preserves diffeomorphism invariance. This approach is known as
multisymplectic field theory, see for example
\cite{Struckmeier:2008zz,Kanatchikov:1997wp,Forger:2002ak}, for review, see \cite{Kastrup:1982qq} and for recent interesting application of this formalism in string theory, see \cite{Lindstrom:2020szt}.


In this short note we  apply covariant Hamiltonian formalism for extended objects in string theory which are fundamental string, p-brane and unstable Dp-brane. These objects have non-linear Lagrangian density and hence it is interesting to see the form of covariant Hamiltonian. We start with the fundamental string action in Nambu-Gotto form. Then we find its covariant canonical formulation and we obtain covariant Hamiltonian density which have remarkably similar form as original Lagrangian density. This is even more striking when we generalize this analysis to the case of $p+1$-dimensional object known as $p-$brane. We derive its covariant Hamiltonian density using two ways. In the first one we start with square-root structure of the action and in the second one we introduce world-volume auxiliary metric. In both cases we obtain covariant Hamiltonian density that has much more complicated
 structure comparing to two dimensional case or to  the standard Hamiltonian formalism where Hamiltonian depends on conjugate momenta quadratically. Finally we consider covariant Hamiltonian analysis for unstable Dp-brane. In this case the analysis is much more complicated due to the presence of the world-volume gauge field. Explicitly, we were not able to invert relation between covariant conjugate momenta and partial derivatives of world-sheet modes in case of standard Dirac-Born-Infeld like action. For that reason we used non-BPS Dp-brane action with auxiliary world-sheet metric which however is not symmetric.
 Note that this is similar procedure how Nambu-Gotto string action is replaced with Polyakov action and it is well known that these two actions are classically equivalent.
 As a result we derive Hamiltonian density for this non-BPS Dp-brane. Then we consider explicit example which is non-BPS D1-brane where we can solve equations of motion for auxiliary metric. As a result we obtain Hamiltonian density that in the limit $V\rightarrow 0$ reduces to the Hamiltonian density for the gas of fundamental string with agreement with the standard Hamiltonian description of tachyon condensation
 \cite{Sen:2000kd,Sen:2003bc}.

Let us outline our results and suggest possible extension of this work. We found Hamiltonian densities in covariant canonical formalism for fundamental string, p-brane and non-BPS Dp-brane. We showed that this form of Hamiltonian density is very complicated in case of p-brane with $p>1$. We also found Hamiltonian density for unstable Dp-brane and we studied tachyon condensation in case of unstable D1-brane and we showed that this result agrees with the vacuum tachyon condensation that corresponds to the gas of fundamental strings
\cite{Sen:2000kd,Sen:2003bc}. On the other hand the fact that covariant Hamiltonian density is non-polynomial in case of higher dimensional p-branes suggests that it is not clear whether it will be useful for the study of higher dimensional objects in string theory. The situation is different in case of the fundamental string where the Hamiltonian density has the same form as corresponding Lagrangian one. This suggests that covariant Hamiltonian formulation of fundamental string could have its meaning in case of fundamental string and should  be studied further. For example, we could analyze T-duality properties of fundamental string in this picture. We hope to return to this problem in future.

This paper is organized as follows. In the next section (\ref{second}) we find covariant Hamiltonian density for fundamental string. Then we generalize this analysis to the case of p-brane in section (\ref{third}). Finally in section (\ref{fourth}) we find covariant Hamiltonian density for unstable Dp-brane and solve an explicit example given by unstable D1-brane.


 \section{Covariant Hamiltonian formalism for fundamental string}\label{second}
 We begin with standard Nambu-Gotto action for fundamental string.
 Let us consider string action
 \begin{equation}
 S=-T\int d^2\sigma \sqrt{-\det g}=\int d^2\sigma \mL \ ,
 \end{equation}
 where
 \begin{equation}
 g_{\alpha\beta}=G_{MN}\partial_\alpha x^M\partial_\beta x^N \ ,
 \end{equation}
 where $\sigma^\alpha \ , \alpha=0,1$ parameterize world-sheet of the string, $G_{MN}$
 is background metric and $x^M, M=0,\dots,25$ parameterize embedding of the string into target space-times. Finally, $T=\frac{1}{l_s^2}$ where $l_s$ is fundamental string length, is string tension.  Then, following  covariant Hamiltonian formalism,
 we introduce conjugate  momenta as
 \begin{equation}\label{palphaM}
 p^\alpha_M=\frac{\partial \mL}{\partial (\partial_\alpha x^M)}=-
 TG_{MN}\partial_\beta x^N g^{\beta\alpha}\sqrt{-\det g} \ ,
 \end{equation}
 where $g^{\alpha\beta}$ is matrix inverse to $g_{\alpha\beta} \ ,
 g_{\alpha\beta}g^{\beta\gamma}=\delta_\alpha^\gamma$.
 Now using (\ref{palphaM}) we define canonical Hamiltonian density as
 \begin{eqnarray}
 & &\mH=p^\alpha_M \partial_\alpha x^M-\mL=\nonumber \\
& & -TG_{MN}\partial_\beta x^N\partial_\alpha x^M g^{\beta\alpha}\sqrt{-\det g}+
 T\sqrt{-\det g}=-T\sqrt{-\det g} \ . \nonumber \\
 \end{eqnarray}
 As the next step we have to express this Hamiltonian density using
 canonical variables. To do this we introduce matrix
 \begin{equation}
 \Pi^{\alpha\beta}=p^\alpha_M G^{MN}p_N^\beta \ .
\end{equation}
Then with the help of (\ref{palphaM}) we find relation between $\Pi^{\alpha\beta}$
and $g_{\alpha\beta}$
 \begin{equation}
 \Pi^{\alpha\beta}=
- T^2 g^{\alpha\gamma}g_{\gamma\omega}g^{\omega\beta}\det g=-T^2 g^{\alpha\beta}\det g
\nonumber \\
 \end{equation}
 so that
 \begin{equation}
 \det \Pi^{\alpha\beta}=T^4 (\det g)^2 \det g^{\alpha\beta}=T^4 \det g
 \end{equation}
 using $\det g^{\alpha\beta}=\frac{1}{\det g}$. Then we obtain covariant  Hamiltonian density for fundamental string in the form
 \begin{equation}\label{mHstring}
 \mH=-\frac{1}{T}\sqrt{-\det \Pi^{\alpha\beta}} \ .
 \end{equation}
 Finally we derive equations of motion for $x^M$ and $p_M^\alpha$
 when we consider canonical form of the action
\begin{equation}
 S=\int d^2\sigma (p_M^\alpha \partial_\alpha x^M-\mH)
 \end{equation}
 and perform its variation with respect to $x^M$ and $p_M^\alpha$
 \begin{eqnarray}
& & \delta S=\int d^2\sigma \left(\delta p_M^\alpha \partial_\alpha x^M+
 p_M^\alpha \partial_\alpha \delta x^M-\frac{\delta \mH}{\delta x^M}
 \delta x^M-\frac{\delta \mH}{\delta p_M^\alpha}\delta p_M^\alpha\right)=\nonumber \\
& & =\int d^2\sigma \left(\left(\partial_\alpha x^M-\frac{\delta \mH}{\delta p^\alpha_M}\right)\delta p_M^\alpha
 -\left(\partial_\alpha p^\alpha_M+\frac{\delta \mH}{\delta x^M}\right)\delta x^M\right)=0
 \nonumber \\
\end{eqnarray}
and we obtain following equations of motion
\begin{equation}
\partial_\alpha x^M=\frac{\delta \mH}{\delta p^\alpha_M} \ , \quad
\partial_\alpha p^\alpha_M=-\frac{\delta \mH}{\delta x^M} \ .
\end{equation}
Now for the Hamiltonian density (\ref{mHstring}) we obtain
\begin{equation}
\partial_\alpha x^M=-\frac{1}{T}G^{MN}p_N^\beta \Pi_{\beta\alpha}
\sqrt{-\det \Pi^{\alpha\beta}} \ , \quad
\partial_\alpha p^\alpha_M=\frac{1}{2T}p^\alpha_K\partial_M G^{KL}p^\beta_L
\Pi_{\beta\alpha}\sqrt{-\det \Pi^{\alpha\beta}} \ ,
\end{equation}
where $\Pi_{\alpha\beta}$ is matrix inverse to $\Pi^{\alpha\beta}$.

\section{Covariant Hamiltonian Formalism for p-brane}\label{third}
In this section we perform generalization of the analysis
presented in previous section to the case of $p+1$ dimensional
object known as p-brane. Well known example of such object is M2-brane
in M-theory.  So that let us  consider action
\begin{equation}\label{Sp}
S=-T_p\int d^{p+1}\xi \sqrt{-\det g} \ ,
\end{equation}
where $T_p$ is $p-$brane tension and where induced metric has the form
\begin{equation}
g_{\alpha\beta}=G_{MN}\partial_\alpha x^M\partial_\beta x^N \ .
\end{equation}
Finally, $\xi^\alpha,\alpha=0,1,\dots,p$ are world-volume coordinates so that
$\partial_\alpha\equiv \frac{\partial}{\partial \xi^\alpha}$.
Then covariant momenta are equal to
\begin{equation}\label{palphaMp}
p^\alpha_M=\frac{\partial \mL}{\partial \partial_\alpha x^M}
=-T_p G_{MN}\partial_\beta x^N g^{\beta\alpha}\sqrt{-\det g}
\end{equation}
and hence Hamiltonian is equal to
\begin{equation}\label{mHgp}
\mH=p^\alpha_M\partial_\alpha x^M-\mL=
-pT_p \sqrt{-\det g} \
\end{equation}
using the fact that $g_{\alpha\beta}g^{\beta\alpha}=p+1$.
We again introduce matrix $\Pi^{\alpha\beta}=p^\alpha_M G^{MN}
p^\beta_N$ and then from (\ref{palphaMp}) we obtain
\begin{equation}
\Pi^{\alpha\beta}=-T_p^2 g^{\alpha\beta}\det g \ .
\end{equation}
Taking determinant of these matrices we obtain relation between $\det g$ and
$\det \Pi^{\alpha\beta}$ in the form
\begin{equation}
-\det g=(-\det \Pi)^{\frac{1}{p}} \ .
\end{equation}
Inserting this result into (\ref{mHgp}) we obtain final form of the covariant
Hamiltonian density for p-brane in the form
\begin{equation}\label{mHp}
\mH=-pT_p\sqrt{(-\det\Pi)^{1/p}} \ .
\end{equation}
We see that it is much more complicated than in case of fundamental string. This fact suggests limitation of the covariant canonical formalism for higher dimensional objects in string theory.
\subsection{Auxiliary metric}
There is an alternative procedure which is based on the formulation of
p-brane with an auxiliary world-volume metric $\gamma_{\alpha\beta}$. Explicitly,
let us  introduce auxiliary metric $\gamma_{\alpha\beta}$ and write an action in the form
\begin{equation}\label{Spauxiliary}
S=-\frac{1}{2}\int d^{p+1}\xi \sqrt{-\gamma}
(\gamma^{\alpha\beta}g_{\beta\alpha}-(p-1)) \ .
\end{equation}
To see an equivalence of this action with the original one let us solve the
equation of motion for  $\gamma_{\alpha\beta}$ that follow from the action
(\ref{Spauxiliary})
\begin{eqnarray}
-\frac{1}{2}\gamma_{\beta\alpha}(\gamma^{\gamma\delta}g_{\delta\gamma}-(p-1))
+g_{\beta\alpha}=0
\nonumber \\
\end{eqnarray}
that has solution
\begin{equation}
\gamma_{\alpha\beta}=g_{\alpha\beta} \ .
\end{equation}
Then it is easy to see that inserting this solution into (\ref{Spauxiliary}) we get
the original action (\ref{Sp}).
Now from the action (\ref{Spauxiliary}) we obtain
conjugate momenta
\begin{eqnarray}
p^\alpha_M=-\sqrt{-\gamma}\gamma^{\alpha\beta}\partial_\beta x^N g_{NM}
\nonumber \\
\end{eqnarray}
so that
\begin{equation}\label{mHgamma}
\mH=-\frac{1}{2\sqrt{-\gamma}}p^\alpha_M p^\beta_N G^{MN}\gamma_{\alpha\beta}-
\frac{1}{2}\sqrt{-\gamma}(p-1) \ .
\end{equation}
Let us now perform variation with respect to $\gamma_{\alpha\beta}$ in (\ref{mHgamma}) and we obtain equations of motion
\begin{equation}\label{eqgamma}
\frac{1}{4\sqrt{-\gamma}}
\gamma^{\alpha\beta}\Pi^{\gamma\delta}\gamma_{\delta\gamma}
-\frac{1}{2\sqrt{-\gamma}}\Pi^{\alpha\beta}-\frac{1}{4}\sqrt{-\gamma}\gamma^{\alpha\beta}(p-1)=0 \ .
\end{equation}
We will presume solution in the form $\gamma^{\alpha\beta}=K\Pi^{\alpha\beta},
\gamma_{\alpha\beta}=\frac{1}{K}\Pi_{\alpha\beta} \ , \Pi_{\alpha\beta}\Pi^{\beta\gamma}=
\delta_\alpha^\gamma$. Inserting this ansatz into (\ref{eqgamma}) we obtain equation for $K$
\begin{equation}
\frac{1}{2}K
(\frac{p+1}{K}+\frac{p-1}{K^{p+1}\det\Pi^{\alpha\beta}})=1
\end{equation}
that has solution
\begin{equation}
K=(-\frac{1}{\det \Pi^{\alpha\beta}})^{1/p}
\ .
\end{equation}
Inserting this result into (\ref{mHgamma}) we obtain Hamiltonian density in the form
\begin{equation}
\mH=-p\sqrt{(-\det \Pi)^{1/p}} \ .
\end{equation}
This agrees with the Hamiltonian density (\ref{mHp}).
\section{Covariant Hamiltonian Formalism for Non-BPS Dp-brane}\label{fourth}
Finally we proceed to the covariant Hamiltonian formalism for non-BPS Dp-brane whose
action has the form
\begin{equation}\label{Snon}
S=-\tau_{p}\int d^{p+1}\xi e^{-\phi}V
\sqrt{-\det \bA_{\alpha\beta}}
\end{equation}
where
\begin{equation}
\bA_{\alpha\beta}=G_{MN}\partial_\alpha x^M\partial_\beta x^N+
B_{MN}\partial_\alpha x^M\partial_\beta x^N+l_s^2\partial_\alpha T\partial_\beta T+
l_s^2F_{\alpha\beta} \ , \quad  F_{\alpha\beta}=\partial_\alpha A_\beta-\partial_\beta A_\alpha
\end{equation}
where $G_{MN},B_{MN}$ are background metric and NSNS two form field, $T$ is tachyon that propagates on the world-volume of non-BPS Dp-brane with the tachyon potential $V(T)$ that has minimum at $T=\pm \infty$ with $V(T_{min})=0$ and maximum at $T=0$ with $V(T=0)=1$. Finally $\tau_p$ is tension of non-BPS Dp-brane and   $\phi$ is background dilaton. In what follows we presume that  $B_{MN}=0$.

The presence of the gauge field $A_{\alpha}$ makes the analysis very complicated since now matrix $\bA_{\alpha\beta}$ is  non-symmetric and hence it
is  difficult to find relation between covariant momenta and derivatives
of world-sheet modes. For that reason we use an alternative formulation of non-BPS Dp-brane
with an auxiliary metric
 $\gamma_{\alpha\beta}$ and write an action in the form
\begin{equation}\label{actexten}
S=-\frac{\tau_p}{2}\int d^{p+1}\xi \sqrt{-\gamma}e^{-\phi}V
(\gamma^{\alpha\beta}\bA_{\beta\alpha}-(p-1)) \ ,
\end{equation}
where $\gamma_{\alpha\beta}$ is non-symmetric.
Now the equation of motion for $\gamma_{\alpha\beta}$ has the form
\begin{equation}
-\frac{1}{2}\gamma_{\beta\alpha}(\gamma^{\gamma\delta}\bA_{\delta\gamma}-(p-1))
+\bA_{\beta\alpha}=0
\end{equation}
that has solution $\gamma_{\alpha\beta}=\bA_{\alpha\beta}$. Inserting this result into
(\ref{actexten}) we obtain an action (\ref{Snon}) that proves an equivalence
of these two formulations.

Let us then start with (\ref{actexten}) in order to formulate covariant canonical
formalism. From (\ref{actexten}) we obtain
\begin{eqnarray}\label{defpi}
& &\pi^{\alpha\beta}=\frac{\partial \mL}{\partial (\partial_\beta A_\alpha)}=-\frac{\tau_pl_s^2}{2}e^{-\phi}V \gamma^{\alpha\beta}_A \sqrt{-\gamma}\ , \nonumber \\
& &p^\alpha_M=\frac{\partial \mL}{\partial (\partial_\alpha x^M)}=-\tau_p e^{-\phi}V\sqrt{-\gamma}\gamma^{\alpha\beta}_S\partial_\beta x^N G_{NM}
\nonumber \\
& &p^\alpha_T=\frac{\partial \mL}{\partial (\partial_\alpha T)}=-\tau_pl_s^2 e^{-\phi}V\sqrt{-\gamma}\gamma^{\alpha\beta}_S\partial_\beta T  \ ,
\nonumber \\
\end{eqnarray}
where $\gamma^{\alpha\beta}_{S}=\frac{1}{2}(\gamma^{\alpha\beta}+\gamma^{\beta\alpha}) \ ,
\gamma^{\alpha\beta}_A=\frac{1}{2}(\gamma^{\alpha\beta}-\gamma^{\beta\alpha})$.
Using these results we get  Hamiltonian density in the form
\begin{eqnarray}
& &\mH=p^\alpha_M\partial_\alpha x^M+p^\alpha_T\partial_\alpha T+\pi^{\alpha\beta}\partial_\beta A_\alpha-\mL=
\nonumber \\
& &=-\frac{\tau_p}{2}e^{-\phi}V\sqrt{-\gamma}\gamma^{\alpha\beta}(g_{\beta\alpha}+l_s^2\partial_\beta T\partial_\alpha T)-
\frac{\tau_p}{2}\sqrt{-\gamma}e^{-\phi}V(p-1) \ . \nonumber \\
\end{eqnarray}
As the next step we have to express Hamiltonian density with the help of canonical
variables. Using the first equation in (\ref{defpi}) we obtain
\begin{equation}
\gamma^{\beta\alpha}=\gamma^{\alpha\beta}+\frac{4}{\tau_pl_s^2 e^{-\phi}V\sqrt{-\gamma}}\pi^{\alpha\beta} \ .
\end{equation}
Then inserting this result into remaining equations in (\ref{defpi}) we obtain
\begin{eqnarray}
& &p^\alpha_M=
- V^{\alpha\beta}\partial_\beta x^N G_{NM} \ ,  \quad p_T^\alpha=-l_s^2V^{\alpha\beta}\partial_\beta T \ ,
\nonumber \\
& & V^{\alpha\beta}=\tau_p e^{-\phi}V\sqrt{-\gamma}\gamma^{\alpha\beta}+2\pi^{\alpha\beta}l_s^{-2} \ .
\nonumber \\
\end{eqnarray}
Let us introduce  inverse matrix $W_{\alpha\beta}V^{\beta\gamma}=\delta_\alpha^\gamma$
so that we can express $\partial_\alpha x^M,\partial_\alpha T$ with the help of canonical variables as
\begin{equation}
\partial_\alpha x^N=-W_{\alpha\beta}p^\beta_M G^{MN} \ , \quad
\partial_\alpha T=-W_{\alpha\beta}p^\beta_T \ .
\end{equation}
As a result we obtain final form of the covariant Hamiltonian density for non-BPS Dp-brane
\begin{eqnarray}
\nonumber \\
\mH=-\frac{1}{2}W_{\alpha\beta}\Pi^{\beta\alpha} -\frac{\tau_p}{2}\sqrt{-\gamma}e^{-\phi}
V(p-1) \ , \quad  \Pi^{\alpha\beta}=p^\alpha_{ \ M}G^{MN}p^\beta_{ \ N} \ . \nonumber \\
\end{eqnarray}
This is final form of covariant Hamiltonian density for non-BPS Dp-brane. Unfortunately we were not able to find explicit form of the matrix $W_{\alpha\beta}$ except of the case of unstable D1-brane.
\subsection{Solvable Example: Non-BPS  D1-brane}
In case of unstable D1-brane it is convenient to write $\gamma^{\alpha\beta}=
s^{\alpha\beta}+a^{\alpha\beta}$ where $a^{\alpha\beta}$ is antisymmetric matrix that can be chosen as
\begin{equation}
a^{\alpha\beta}=A\epsilon^{\alpha\beta} \ , \quad \epsilon_{\alpha\beta}=
\left(\begin{array}{cc}
0 & -1 \\
1 & 0 \\ \end{array}\right) \ , \quad
\epsilon^{\alpha\beta}=\left(\begin{array}{cc}
0 & 1 \\
-1 & 0 \\ \end{array}\right) \ , \quad  \epsilon_{\alpha\beta}\epsilon^{\beta\gamma}=\delta_\alpha^\gamma \ .
\end{equation}
Then momentum conjugate to $A_\alpha$ is equal to
\begin{equation}
\pi^{\alpha\beta}=-\frac{l_s^2\tau_1}{2}e^{-\phi}VA\epsilon^{\alpha\beta}\frac{1}{\sqrt{-\det s^{\alpha\beta}-A^2}}
	\end{equation}
	that implies
\begin{equation}\label{A}
A=
\frac{l_s^{-2}\pi \sqrt{-\det s^{\alpha\beta}}}{\sqrt{l_s^{-4}\pi^2+
\tau_1^2 e^{-2\phi}V^2}} \ , \quad \pi\equiv \pi^{\alpha\beta}\epsilon_{\beta\alpha} \ .
\end{equation}	
Further, $p^\alpha_M$ and $p_T^\alpha$ are equal to
\begin{eqnarray}
& &p^\alpha_M=-\tau_1 e^{-\phi}V G_{MN}\partial_\beta x^N s^{\beta\alpha}
\frac{1}{\sqrt{-\det s^{\alpha\beta}-A^2}} \ , \nonumber \\
& & p^\alpha_T=-\tau_1 l_s^2e^{-\phi}V \partial_\beta T s^{\beta\alpha}
\frac{1}{\sqrt{-\det s^{\alpha\beta}-A^2}} \ . \nonumber \\
\end{eqnarray}
Then using $A$ given in (\ref{A}) we get final result
\begin{equation}
\mH=-\frac{\sqrt{-\det s^{\alpha\beta}}}{2\sqrt{l_s^{-4}\pi^2+\tau_1^2 e^{-2\phi}V^2}}s_{\alpha\beta}(\Pi^{\alpha\beta}+l_s^{-2}p_T^\alpha p_T^\beta) \ .
\end{equation}	
Finally we solve equation of motion for $s_{\alpha\beta}$
\begin{equation}
\frac{1}{2\sqrt{-\det s_{\alpha\beta}}}s^{\alpha\beta}s_{\gamma\delta}
(\Pi^{\gamma\delta}+l_s^{-2}p_T^\gamma p_T^\delta)
-\frac{1}{\sqrt{-\det s}}
(\Pi^{\alpha\beta}+l_s^{-2}p_T^\alpha p_T^\beta)
=0
\end{equation}
that can be solved as $s^{\alpha\beta}=\Pi^{\alpha\beta}+l_s^{-2}p_T^\alpha p_T^\beta $ so that Hamiltonian density has the form
\begin{equation}\label{HconD1}
\mH=-\frac{1}{\sqrt{l_s^{-4}\pi^2+\tau_1^2 e^{-2\phi}V^2}}
\sqrt{-\det (\Pi^{\alpha\beta}+l_s^{-2}p_T^\alpha p_T^\beta)} \ .
\end{equation}
This is final form of the covariant Hamiltonian density for unstable D1-brane.
Now we see that when the tachyon reaches its minimum everywhere on the world-volume
of D1-brane we have $T=T_{min},p_T^\alpha=0$ where $V(T_{min})=0, \frac{dV}{dT}(T_{min})=0$, the
Hamiltonian density (\ref{HconD1}) takes the form
\begin{equation}
\mH=-\frac{l_s^2}{|\pi|}\sqrt{-\det \Pi^{\alpha\beta}}
\end{equation}
which is a covariant density for the bound state of $|\pi|$ fundamental strings which is nice confirmation of the tachyon vacuum condensation \cite{Sen:2000kd,Sen:2003bc}.


\begin{thebibliography}{20}

\bibitem{Polchinski:1998rq}
J.~Polchinski,
\emph{``String theory. Vol. 1: An introduction to the bosonic string,''}
doi:10.1017/CBO9780511816079


\bibitem{Polchinski:1998rr}
J.~Polchinski,
\emph{``String theory. Vol. 2: Superstring theory and beyond,''}
doi:10.1017/CBO9780511618123

\bibitem{Ortin:2015hya}
T.~Ortin,
\emph{``Gravity and Strings,''}
doi:10.1017/CBO9781139019750

\bibitem{Sen:2004nf}
A.~Sen,
\emph{``Tachyon dynamics in open string theory,''}
Int. J. Mod. Phys. A \textbf{20} (2005), 5513-5656
doi:10.1142/S0217751X0502519X
[arXiv:hep-th/0410103 [hep-th]].


\bibitem{Sen:1999md}
A.~Sen,
\emph{``Supersymmetric world volume action for nonBPS D-branes,''}
JHEP \textbf{10} (1999), 008
doi:10.1088/1126-6708/1999/10/008
[arXiv:hep-th/9909062 [hep-th]].

\bibitem{Bergshoeff:2000dq}
E.~Bergshoeff, M.~de Roo, T.~de Wit, E.~Eyras and S.~Panda,
\emph{``T duality and actions for nonBPS D-branes,''}
JHEP \textbf{05} (2000), 009
doi:10.1088/1126-6708/2000/05/009
[arXiv:hep-th/0003221 [hep-th]].

\bibitem{Kluson:2000iy}
J.~Kluson,
\emph{``Proposal for nonBPS D-brane action,''}
Phys. Rev. D \textbf{62} (2000), 126003
doi:10.1103/PhysRevD.62.126003
[arXiv:hep-th/0004106 [hep-th]].





\bibitem{Sen:2000kd}
A.~Sen,
\emph{``Fundamental strings in open string theory at the tachyonic vacuum,''}
J. Math. Phys. \textbf{42} (2001), 2844-2853
doi:10.1063/1.1377037
[arXiv:hep-th/0010240 [hep-th]].


\bibitem{Sen:2003bc}
A.~Sen,
\emph{``Open and closed strings from unstable D-branes,''}
Phys. Rev. D \textbf{68} (2003), 106003
doi:10.1103/PhysRevD.68.106003
[arXiv:hep-th/0305011 [hep-th]].

\bibitem{DeDonder}
Th. De Donder, \emph{"Théorie Invariantive Du Calcul des Variations"},
 (Gaulthier-Villars and Cie., Paris, 1930)

\bibitem{Weyl}
H. Weyl, \emph{"Geodesic Fields in the Calculus of Variation for Multiple Integrals"}
Annals of Mathematics, {\bf 36} , p.607



\bibitem{Struckmeier:2008zz}
J.~Struckmeier and A.~Redelbach,
\emph{``Covariant Hamiltonian field theory,''}
Int. J. Mod. Phys. E \textbf{17} (2008), 435-491
doi:10.1142/S0218301308009458
[arXiv:0811.0508 [math-ph]].

\bibitem{Kanatchikov:1997wp}
I.~V.~Kanatchikov,
\emph{``Canonical structure of classical field theory in the polymomentum phase space,''}
Rept. Math. Phys. \textbf{41} (1998), 49-90
doi:10.1016/S0034-4877(98)80182-1
[arXiv:hep-th/9709229 [hep-th]].


\bibitem{Forger:2002ak}
M.~Forger, C.~Paufler and H.~Roemer,
\emph{``The Poisson bracket for Poisson forms in multisymplectic field theory,''}
Rev. Math. Phys. \textbf{15} (2003), 705-744
doi:10.1142/S0129055X03001734
[arXiv:math-ph/0202043 [math-ph]].


\bibitem{Kastrup:1982qq}
H.~Kastrup,
\emph{``Canonical Theories of Dynamical Systems in Physics,''}
Phys. Rept. \textbf{101} (1983), 1
doi:10.1016/0370-1573(83)90037-6

\bibitem{Lindstrom:2020szt}
U.~Lindström,
\emph{``Covariant Hamiltonians, sigma models and supersymmetry,''}
[arXiv:2004.01073 [hep-th]].

\end{thebibliography}
\end{document}